# Topological Timing of Bound States by Spacetime Mass Windings

Hamdi Barkaoui

Ministry of Industry and Information Technology Key Lab of Micro-Nano Optoelectronic Information System, Guangdong Provincial Key Laboratory of Semiconductor Optoelectronic Materials and Intelligent Photonic Systems, Harbin Institute of Technology, Shenzhen 518055, P. R. China.

Corresponding email: hamdi@hit.edu.cn

Topology usually determines where a wave is trapped. Here we show that it can also determine when a trapped state brightens or dims. We consider a dimerized photonic lattice in which a real Dirac mass, varying in space, binds a Jackiw-Rebbi state, while staggered gain and loss, switched in time, act as an imaginary Dirac mass on that same state. Together, the two masses form a mass field in space and time whose isolated zeros carry an integer winding. Combining a continuum Dirac description with finite-lattice simulations, we show that the sign of this winding fixes whether the bound state reaches a maximum or a minimum at the switch, while the gain and loss remain below the bulk gap. The windings obey a conservation law: they are created and annihilated in pairs, and their sum along a boundary determines which extrema are protected. Full-wave simulations of a passive waveguide array, in which the propagation distance plays the role of time, reproduce these predictions with loss alone. These results provide a topological handle on the timing of localized optical responses, relevant to switching, gating, and pulse shaping of confined modes in integrated photonic circuits and topological lasers.

## 1- Introduction

Topology is most often used to decide where a wave is trapped. In Dirac systems, a mass that changes sign binds a state to the domain wall, the Jackiw-Rebbi (JR) mechanism[1], and the Su-Schrieffer-Heeger (SSH) lattice provides its simplest realization, with the dimerization playing the role of the mass[2]. This principle underlies much of topological photonics, where interfaces between distinct phases fix the position of optical states[3–6]. SSH interface states have been realized in coupled resonators, waveguide arrays, and semiconductor lasers, where they serve as robust localized modes for light confinement and single-mode lasing[7,8].

Time has recently joined space as a stage for topology. Photonic time crystals and temporal interfaces[9–12] support states in momentum gaps[13–15], and combined modulations can localize light in both space and time[16–18]. Gain and loss, which enter the Dirac description as anti-Hermitian masses[19], play a central role: reversing the gain-loss profile of a crystal in its PT-broken phase binds a temporal analogue of the JR state[20], and quantum walks with both energy and momentum gaps host spacetime-topological events[17]. In all these cases, however, the temporal topology belongs to the bulk bands: it requires a momentum gap, and hence an amplifying bulk.

A different question arises for states that are already trapped: can topology decide when such a state brightens or dims? When gain and loss remain weak compared with the bulk gap, the bulk spectrum stays real and no band invariant changes in time. Yet weak staggered gain and loss still act strongly on a bound state, because they couple selectively to its sublattice polarization[21,22]. Temporal modulation of existing optical states is widely used for switching, frequency conversion, nonreciprocity, and amplification[23,24], and a topological description of such control would make its outcome robust to the details of the modulation. What has been missing is a topological quantity, defined locally in space and time, that predicts this response.

In this Letter, we show that such a quantity arises when two Dirac masses of different character are combined. A real mass, varying in space, binds a JR state; an imaginary mass, varying in time, acts on that same state. Together they form a mass field in space and time whose isolated zeros carry an integer winding. Its sign decides whether the trapped state peaks or dips when the imaginary mass changes sign, and hence the timing of its response. The windings also obey a conservation law: they are created and annihilated in pairs, and their sum on a boundary fixes which extrema are protected. We first formulate this local topology and test it in non-Hermitian SSH chains, including nonseparable mass fields and multiple zeros. We then show that the effect survives in full-wave simulations of a passive waveguide array, in which the propagation distance plays the role of time, and discuss how it can be observed experimentally.

## 2- Principles and Methods

We consider an SSH lattice whose dimerization varies in space, while its two sublattices experience opposite, time-dependent gain and loss (see Fig. 1(a)). Near the band inversion, the lattice is described by the Dirac Hamiltonian[25]

$$H_{\mathrm{D}}(x,t) = \omega_0 I + iv\sigma_y\,\partial_x + m(x)\sigma_x + i\gamma(t)\sigma_z \qquad (1).$$

Here $\omega_0$ is the reference frequency, $v$ is the Dirac velocity, $m$ is the real mass set by the difference between the intracell and intercell couplings, and $\gamma$ is a staggered imaginary mass: sublattices A and B acquire onsite terms $+i\gamma$ and $-i\gamma$. The kinetic term and the two masses are represented by mutually anticommuting Pauli matrices, so that $m$ and $\gamma$ act as two independent components of the Dirac mass. The real mass changes sign at the domain wall $x_d$ and binds a JR state[26,27]; the imaginary mass changes sign at the switching time $t_k$. Throughout, $|\gamma|$ remains smaller than the bulk gap. The bulk frequencies, $\omega_0 \pm [|h(k)|^2 - \gamma^2]^{1/2}$, where $h(k)$ is the coupling term of the Bloch Hamiltonian, then stay real for all Bloch momenta $k$: no momentum gap opens, and the bulk topology does not change at $t_k$ (see Section S1 in the Supplementary Material). The switch therefore cannot create a temporal boundary state of the type discussed above, and only states already bound by $m$ respond to it.

The two masses can be followed together through the mass field $M(x,t) = m(x) + i\gamma(t)$, a compact way of tracking the direction of the vector $(m,\gamma)$. $M$ is not a complex mass entering the Hamiltonian, $m$ and $\gamma$ remain independent coefficients of anticommuting matrices. $M$ vanishes at $(x_d, t_k)$, where the zero lines of $m$ and $\gamma$ intersect (see Fig. 1(b)). A closed counterclockwise contour $C$ around this isolated zero defines the spacetime winding

$$Q_{\mathrm{st}} = \frac{1}{2\pi} \oint_C d \ \arg M \qquad (2),$$

which counts how many times the direction of $(m, \gamma)$ turns around the origin along $C$. Unlike Bloch-band invariants[28,29], $Q_{\mathrm{st}}$ is local: the contour can be drawn around a single zero, so that different zeros of the same system can carry different windings. As for other topological defects[30], $Q_{\mathrm{st}}$ cannot change under smooth deformations of the mass field, unless a zero crosses the contour or two zeros of opposite winding merge. For masses that vary separately in space and time,

$$Q_{\mathrm{st}} = \mathrm{sgn}(\partial_x m \ \partial_t \gamma) \qquad (3)$$

at the zero. In Fig. 1(b), $m$ increases with $x$ and $\gamma$ increases with $t$, so that $\arg M$ increases by $2\pi$ along $C$, and $Q_{\mathrm{st}} = +1$; reversing either kink reverses $Q_{\mathrm{st}}$.

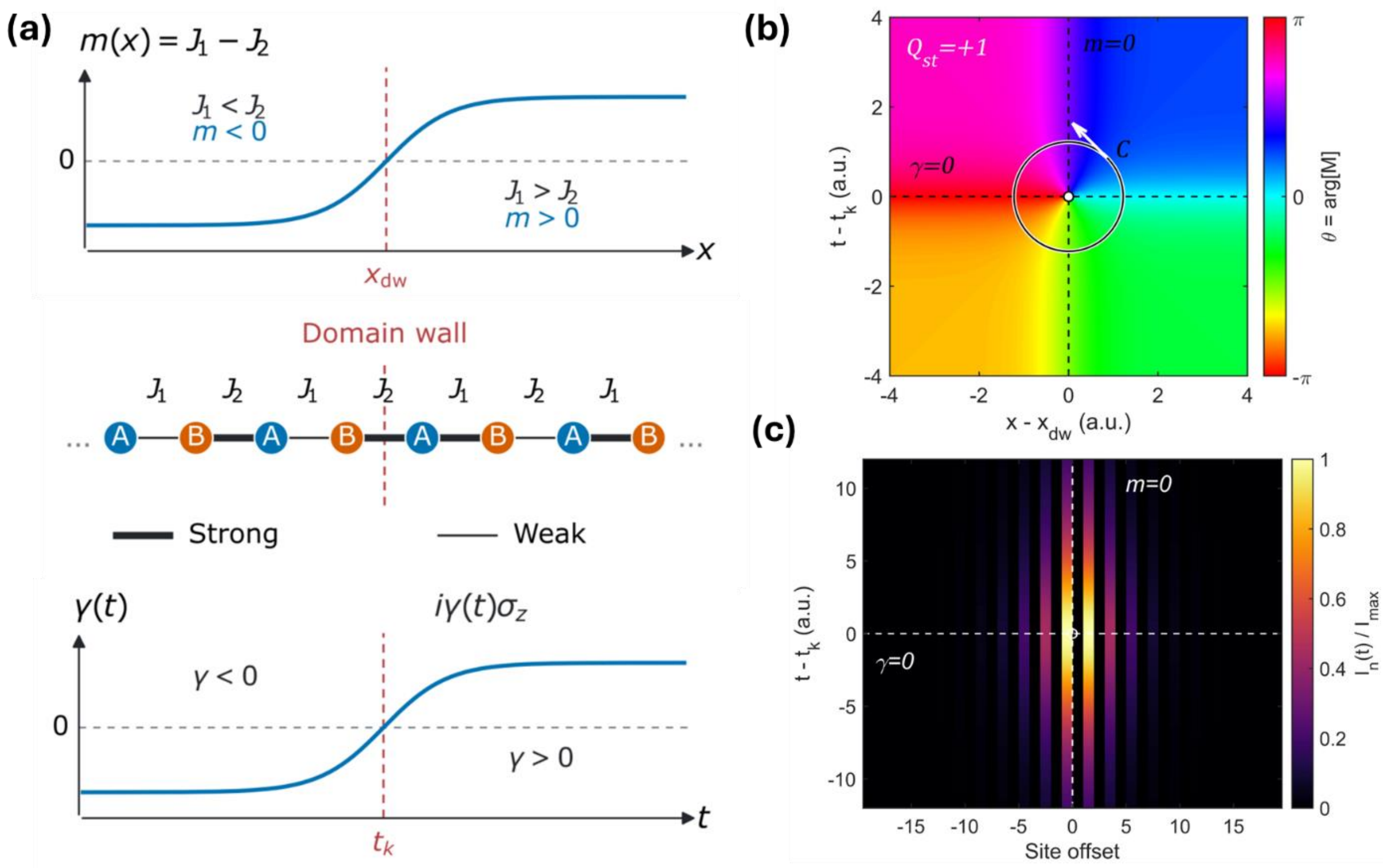

**Figure 1.** (a) SSH lattice with a dimerization kink, giving a real mass $m(x)$ that changes sign at the domain wall $x_{\mathrm{d}}$, and staggered gain and loss, giving an imaginary mass $\gamma(t)$ that changes sign at $t_k$. (b) Phase $\arg M$ of the mass field $M = m + i\gamma$. The zero lines of $m$ and $\gamma$ (dashed) intersect at an isolated zero with winding $Q_{\mathrm{st}} = +1$. (c) Site intensity in a finite chain: the bound state stays at the wall and peaks at $t_k$. Dashed lines: zeros of $m$ and $\gamma$.

The winding acquires a dynamical meaning through the JR state. In the chiral limit, this state occupies a single sublattice, with a polarization $s = -\mathrm{sgn}\,\partial_x m$ set by the orientation of the wall[21]. reversing the wall exchanges the occupied sublattice. Staggered gain and loss therefore act on the bound state as a single rate. Projecting the dynamics onto the JR mode gives its power

$$P(t) = P(t_0)e^{2s\int_{t_0}^{t}\gamma(t')dt'} \qquad (4),$$

so that the state grows when $s\gamma > 0$ and decays when $s\gamma < 0$. This exponential factor has the same form as the envelope of the temporal JR state[20], but its sign is now fixed by the polarization of a spatially bound mode. Because $\gamma$ changes sign at $t_k$, $\ln P$ is extremal there, and its curvature obeys

$$Q_{\mathrm{st}} = -\mathrm{sgn}\left[\partial_t^2 \ln P(t)|_{t_k}\right] \qquad (5).$$

The bound state therefore peaks at $t_k$ for $Q_{\mathrm{st}} = +1$ and dips for $Q_{\mathrm{st}} = -1$. For separable masses, this rule has a simple reading: the state experiences gain before the switch and loss after it, or the reverse. More generally, $Q_{\mathrm{st}}$ is the oriented crossing number of the zero lines of $m$ and $\gamma$, and Eq. (5) is the invariant form of this picture.

The lattice results are obtained for open SSH chains with sites $A_j$ and $B_j$, local mass $m_j = J_{1,j} - J_{2,j}$, and onsite terms $+i\gamma_j(t)$ and $-i\gamma_j(t)$ on the two sublattices. The amplitudes are evolved with an adaptive Runge-Kutta solver, and the power of a bound state is obtained by projecting the evolving field onto it. In a finite chain, the boundary and interface states are nearly degenerate, so that a direct diagonalization can mix them; we separate them by diagonalizing the position operator within the two-dimensional subspace of midgap states. The full-wave simulations of Sec. 3 solve Maxwell's equations for a two-dimensional array of slab waveguides with the beam-envelope method of COMSOL Multiphysics. The parameters of all simulations, the disorder analysis, and further details of the full-wave model are given in Secs. S2 to S4 of the Supplementary Material.

## 3- Results

We now test the response rule of Eq. (5) in finite SSH chains, simulated as described in Sec. 2. In the simplest setting, a short source pulse prepares the bound state well before the switch. The intensity then follows the predicted sequence as shown in Fig. 1(c): it stays at the wall, rises toward $t_k$, peaks there, and decays afterward. Figure 2 examines the mechanism behind this response and isolates the role of the winding. The chain hosts two midgap states, one at its outer boundary and one at the domain wall (see Fig. 2(a)). Separated as described above, they occupy opposite sublattices, with polarizations

$$s_\alpha = \langle\psi_\alpha|\Sigma_z|\psi_\alpha\rangle \qquad (6),$$

where $\Sigma_z = +1$ on A sites and $-1$ on B sites. Equation (6) yields to $s_{\mathrm{b}} = +1$ and $s_{\mathrm{int}} = -1$. A static staggered gain and loss therefore shifts their imaginary frequencies in opposite directions, $\Delta\mathrm{Im}\,\omega_\alpha \simeq s_\alpha\gamma$ as confirmed by the finite-chain eigenvalues in Fig. 2(b). This selective action is known from complex SSH lattices[21], and it is the channel through which the time-dependent imaginary mass acts on the bound state. A gain or loss shared equally by both sublattices would instead shift both states in the same direction and could not distinguish them. The opposite responses of the boundary and interface states persist under simultaneous disorder of up to 30% in the couplings and in the gain-loss amplitudes (Sec. S3 and Fig. S1).

We now keep the interface state fixed and reverse only the temporal kink, $\gamma_{\pm}(t) = \pm\gamma_0 \tanh\,(t - t_k/\tau\,)$, where $\gamma_0$ is the amplitude of the imaginary mass and $\tau$ the switching time. This reverses $Q_{\rm st}$ without changing where the state lives or which sublattice it occupies (see Fig. 2(c)). For $Q_{\rm st} = +1$, the state grows before $t_k$ and decays afterward, producing a maximum; for $Q_{\rm st} = -1$, the sequence is reversed and a minimum appears at the same time. The full-chain dynamics closely follow the single mode prediction of Eq. (4), so the reversal involves neither bulk modes nor a change of the bound state.

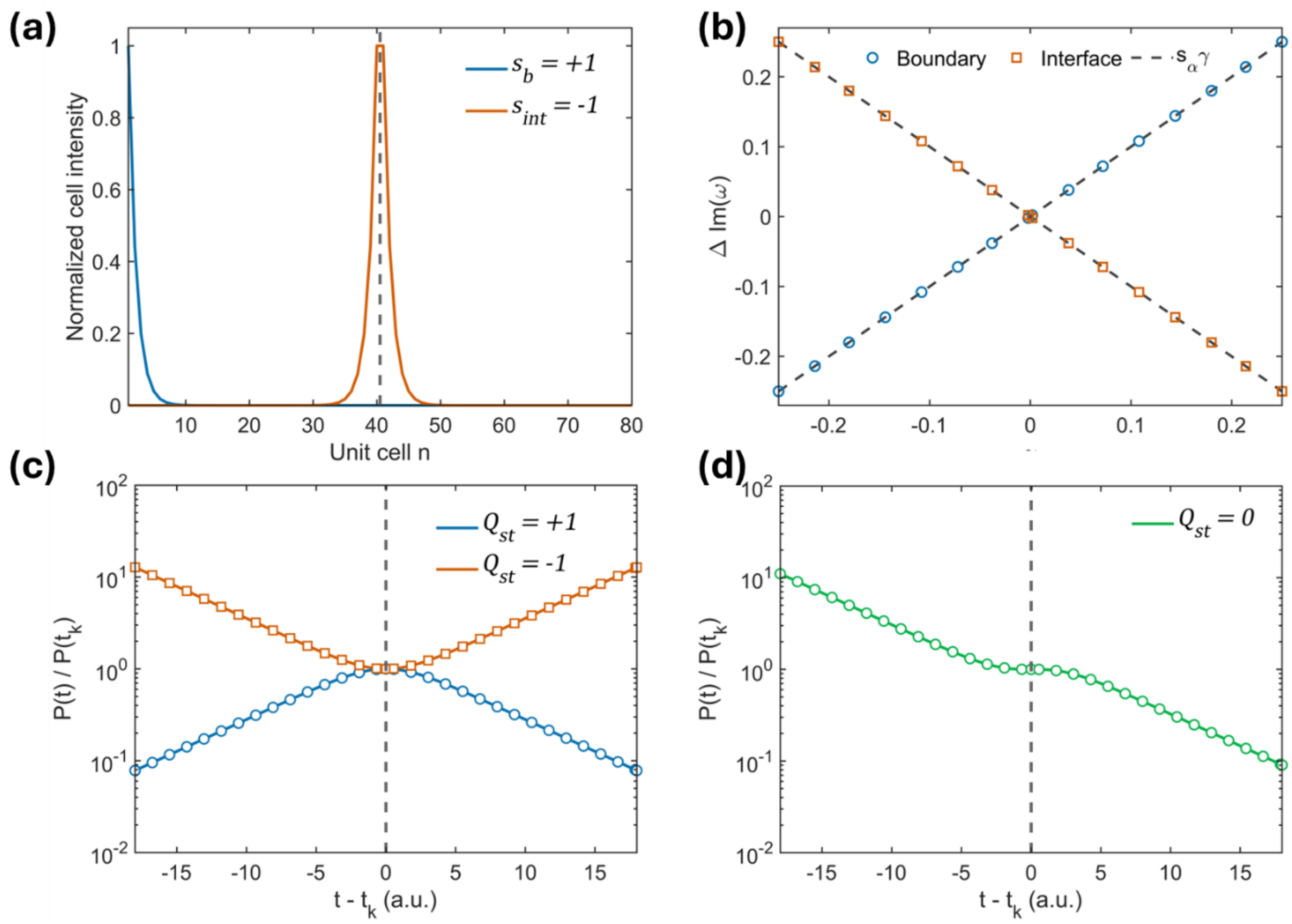

**Figure 2.** (a) Passive boundary and interface states of a finite chain, each normalized to its maximum; dotted line: domain wall. (b) Imaginary frequency shifts under a static staggered $\gamma$: finite chain (symbols) and single mode prediction $s_\alpha\gamma$ (dashed lines). (c) Interface power $P(t)/P(t_k)$ for $Q_{\rm st} = +1$ (blue) and $Q_{\rm st} = -1$ (orange): full chain (symbols) and single mode prediction (lines). (d) Zero winding control, $\gamma \propto \tanh^{\,2}$, which produces no extremum. Dashed lines in (c) and (d): $t_k$.

This test alone cannot separate the winding from a simple reversal of gain and loss, since reversing the kink also exchanges the amplifying and attenuating intervals. We therefore remove the winding while keeping the zero. With $\gamma(t) = \gamma_0 \tanh^2(t - t_k/\tau)$, both masses still vanish at $(x_d, t_k)$, but γ does not change sign, the direction of $(m, \gamma)$ no longer turns around the origin, and $Q_{\mathrm{st}} = 0$ (Sec. S1). The power then becomes stationary at $t_k$ but keeps decreasing, with neither a maximum nor a minimum (see Fig. 2(d)). A common zero of the two masses is therefore not enough: the extremum requires a nonzero winding.

So far, the two masses varied independently, one in space and one in time. One may then ask whether the winding is more than a label for a static interface followed by a separate gain-loss switch. To test this, we let both masses depend on space and time. The zero lines of $m$ and $\gamma$ are then no longer vertical and horizontal in the $(x, \mathrm{t})$ plane: the wall can move along a trajectory $x_{w(t)}$ defined by $m(x_w, t) = 0$ and the switch can reach different parts of the chain at different times. The two zero lines still cross at isolated zeros, whose winding is set by the relative orientation of the two mass gradients:

$$Q_{\mathrm{st}} = \mathrm{sgn}\mathcal{J} = \mathrm{sgn}[\partial_x m \ \partial_t \gamma - \partial_t m \ \partial_x \gamma] \qquad (7).$$

For separable masses, the cross derivatives vanish and Eq. (7) reduces to Eq (3). The meaning of Eq. (7) follows from the bound state itself. A state that follows the moving wall adiabatically experiences the imaginary mass $\gamma_w(t) = \gamma(x_w(t), t)$ along its trajectory, and at the zero this mass changes sign at the rate $\mathrm{d}\gamma_w(t)/\mathrm{dt} = \mathcal{J}/\,\partial_x m$. Combined with the polarization $s = -\mathrm{sgn}[\partial_x m]$ of the state, this gives back the rule of Eq. (5), now with the general winding. Equation (7) applies to simple zeros, where the zero lines of $m$ and $\gamma$ cross at a nonzero angle. At a degenerate zero, where the two lines touch tangentially, $\mathcal{J}$ vanishes and the bound state need not develop an extremum; we return to this case below.

It is possible to test these predictions using masses that mix space and time, $m = m_0 \tanh(X - aT)$ and $\gamma = \gamma_0 \tanh(T - bX)$, where $X$ and $T$ are the distance from the wall and the time from the switch, in units of the kink widths. As shown in Fig. 3(a), the wall now drifts across the chain, and the zero line of $\gamma$ is tilted, so that the switch reaches different cells at different times. Equation (7) gives $Q_{\mathrm{st}} = \mathrm{sgn}[1 - ab]$. The bound state follows the drifting wall without spreading into the bulk, and along its path it experiences $\gamma_w = \gamma_0 \tanh[(1 - ab)T]$. For $a = 0.25$, the choices $b = 2$ and $b = 6$ therefore give opposite windings, and the state reaches a maximum and a minimum, respectively [Fig. 3(a), inset]. This example is instructive because the two configurations look alike locally: in both, $m$ increases across the wall and $\gamma$ increases in time at any fixed position, so the separable formula would predict a maximum in both cases. Only the tilt of the zero line of $\gamma$, through the cross term of Eq. (7), distinguishes them. The response is thus set by the local orientation of the full mass field rather than by a separate switch.

The locality of the winding appears when several zeros coexist. In Fig. 3(b), a single temporal switch crosses two domain walls of opposite orientation (dashed yellow lines), producing zeros with $Q_{\mathrm{st}} = -1$ and $+1$. Under the same drive, the left state dips at $t_k$ while the right state peaks: each state follows its own zero, not the global modulation. We then reverse the sign of $\gamma$ inside a small region of the $(x, \mathrm{t})$ plane (dotted ellipse). In a waveguide array, such a region would simply be a patch of guides in which the loss assignment is exchanged over a finite propagation length. Its boundary is a closed zero line of $\gamma$, which crosses the right wall twice and creates two additional zeros of opposite winding. The right state responds exactly as predicted: after its maximum at $t_k$, it dips as it enters the region and peaks again as it leaves. The windings fix the sequence of extrema, while their heights reflect the gain and loss accumulated in between. The second maximum, for instance, is higher than the first, because the gain inside the reversed region acts at full strength over a longer interval than the loss that precedes it, which only builds up gradually after the switch.

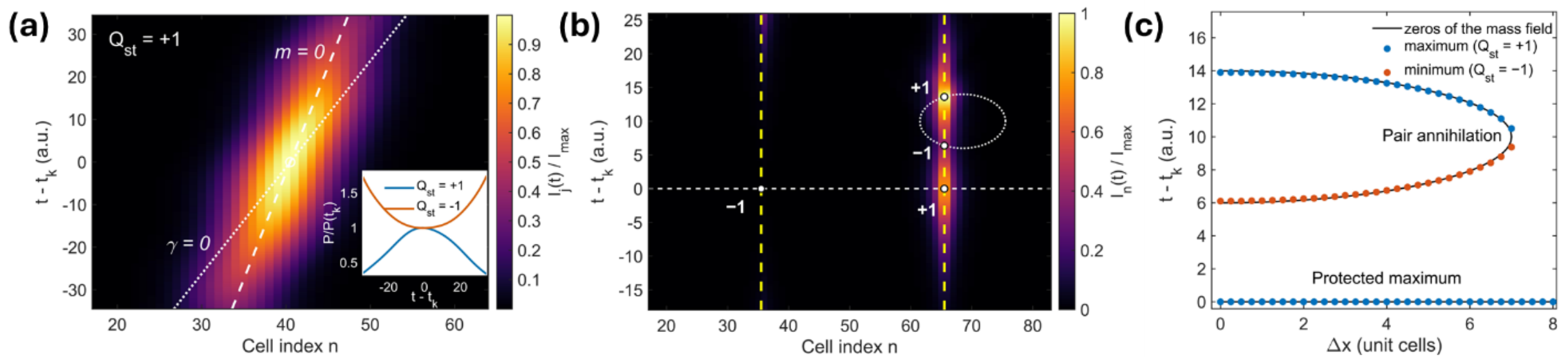

**Figure 3.** (a) Bound state in a nonseparable mass field: the domain wall (dashed line) drifts in time and crosses the zero line of γ (dotted line) at a zero with $Q_{\mathrm{st}} = +1$. Inset: power for $Q_{\mathrm{st}} = \pm 1$, showing opposite extrema. (b) Two domain walls of opposite orientation (dashed yellow lines) under one temporal switch, with $\gamma$ reversed inside the dotted ellipse, displaced by $\Delta x$ from the right wall. Circles: zeros of the mass field, labeled by their windings. (c) Times of the extrema of the right bound state (dots) versus $\Delta x$, compared with the predicted zeros (line). The pair created by the reversed region annihilates at $\Delta x = 7$ cells, the half-width of the region, while the maximum at $t_k$ persists.

Because the windings are integers, they also obey a global rule. Evaluating Eq. (2) along the boundary of any region R of the $(x, \mathrm{t})$ plane gives a boundary winding $W_{\mathrm{R}}$, which counts the turns of $(m, \gamma)$ along that boundary. Since the direction of $(m, \gamma)$ varies smoothly wherever $M \neq 0$, the boundary can be shrunk onto small circles around the zeros it encloses, so that $W_{\mathrm{R}}$ equals the sum of their windings; $Q_{\mathrm{st}}$ is the special case of a region containing a single zero. Denoting by $N_{\mathrm{max}}$ and $N_{\mathrm{min}}$ the numbers of zeros in R at which the corresponding bound state reaches a maximum and a minimum, Eq. (5) gives, for simple and well separated zeros,

$$W_{\mathrm{R}} = N_{\mathrm{max}} - N_{\mathrm{min}} \qquad (8).$$

We can observe in Figure 3(c) this conservation law at work. As the reversed region is moved away from the wall by a distance $\Delta x$, its two zeros approach each other, and the minimum and the extra maximum of the bound state follow them closely. At $\Delta x = 7$ cells, the boundary of the region becomes tangent to the wall: the two zeros merge into a degenerate zero and annihilate, and both extrema disappear together, much like the polarization charges of bound states in the continuum[31,32]. On the other hand, the maximum at $t_k$ is never affected. For a region enclosing the right wall, $W_{\mathrm{R}} = +1$, so this maximum is required by the boundary and cannot be removed by any rearrangement inside it. A region enclosing both walls has instead $W_{\mathrm{R}} = 0$: the minimum of the left state and the maximum of the right state are then allowed but not required, and they would disappear together if the two walls were brought into contact. The boundary thus determines not only how many extrema appear, but which of them are protected. Unintended walls or local defects in the gain-loss pattern can only add maxima and minima in pairs, and cannot remove the extrema required by the boundary.

Finally, we ask whether the effect survives in a real optical structure, where the fields are continuous, every waveguide couples to more than its nearest neighbors, and only loss is readily available. Arrays of evanescently coupled waveguides offer a clean test[33,34]. As light propagates along the array, the propagation distance z plays the role of time, so a temporal kink of the imaginary mass becomes a pattern written into the structure rather than a modulation applied at optical frequencies. The whole spacetime mass field is then frozen into the sample: the domain wall is a fixed arrangement of gaps across the array, and the switch is a gradual change of loss along it. Such arrays have already revealed SSH interface states and loss driven topological transitions[35].

We consider an array of $60$ single-mode slab waveguides with a weak index contrast typical of femtosecond written glass waveguides, at a wavelength of $633\ nm$ (Sec. S4). The gaps between neighboring guides (named A and B) alternate between two values, giving strong and weak couplings of $0.6\ mm^{-1}$ and $0.4\ mm^{-1}$. Exchanging the order of the two gaps at the center of the array creates the domain wall (see Fig. 4(a)). Its bound state lives on a single sublattice, the B guides, with amplitudes that alternate in sign and decrease by a factor of $2/3$ per unit cell away from the wall, as in the finite chain.

The imaginary mass is realized with loss alone. Along z, the loss is gradually transferred from the A guides to the B guides over about ten millimeters around $z_k$, so that the difference between the losses of the two sublattices changes sign at $z_k$. In coupled-mode form, the guide amplitudes $\psi$ obey

$$i\,\frac{d\psi}{dz} = [H - i\Gamma + i\Gamma g(z)\,\sigma_z]\,\psi \qquad (9),$$

where $H$ contains the passive couplings, $\sigma_z = +1$ on the A guides and $-1$ on the B guides, and $g(z)$ describes the loss transfer, with $g(z) = \tanh[(z - z_k)/\tau]$ for $Q_{\mathrm{st}} = +1$. The last term is the staggered imaginary mass of Eq. (1), with $z$ in place of $t$ and $\gamma = \Gamma g$. The term $-i\Gamma$ is a loss shared equally by all guides, which is the only difference between this passive structure and one with balanced gain and loss. It multiplies every amplitude by $e^{-\Gamma z}$, and therefore adds a straight line to $\ln P$ without changing its curvature: it dims the light uniformly but cannot create or remove an extremum. We therefore display the intensities multiplied by $e^{2\Gamma z}$. The mean loss rate, $\Gamma = 0.025\ mm^{-1}$, corresponds to $0.43\ dB/mm$ in a fully lossy guide, comparable to the loss already demonstrated in such arrays [35]. The resulting imaginary mass remains well below the bulk gap set by the difference of the two couplings, so that, as in the lattice model, the switch acts on the bound state without changing the bulk topology.

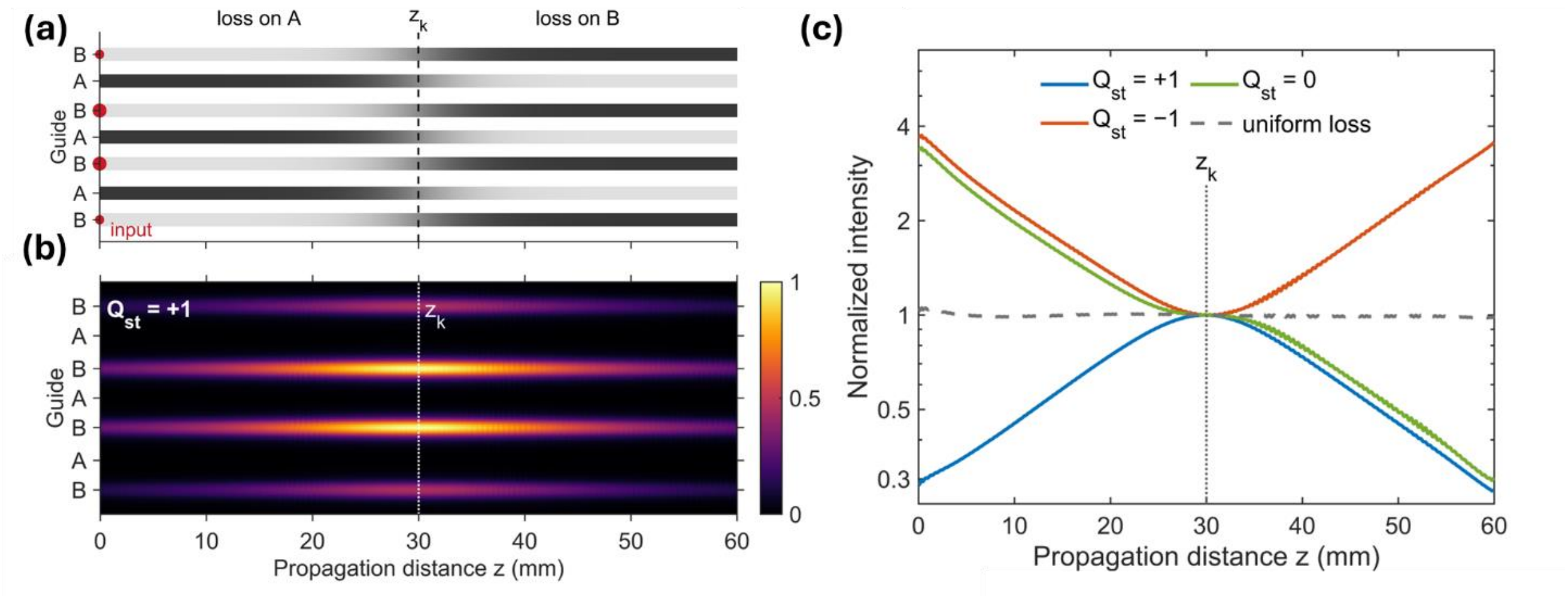

**Figure 4.** (a) Waveguides near the domain wall of a passive array, in which the propagation distance $z$ plays the role of time. Grey level: loss (dark = lossy), switched from the A to the B guides at $z_k$ ($Q_{\mathrm{st}} = +1$). Red dots: input bound state. (b) Intensity $|E|^2 e^{2\Gamma z}$ for $Q_{\mathrm{st}} = +1$, where the factor $e^{2\Gamma z}$ removes the loss rate $\Gamma$ common to all guides. (c) Bound-state intensity $Pe^{2\Gamma z}$, normalized at $z_k$, for $Q_{\mathrm{st}} = \pm 1$ and two controls without winding.

We solve Maxwell equations for the full array with the beam-envelope method, launching the bound state at the input facet (Sec. S4). Full-wave simulations confirm both roles of the mass field. When the loss moves across the sublattices, the light stays on the guides that carry the bound state as displayed in Fig. 4(b). By contrast, the other sublattice remains dark. Indeed, the switch changes how bright the state is, not where it lives. Its brightness, in turn, follows the winding as shown in Fig. 4(c). For $Q_{\mathrm{st}} = +1$, the B guides are almost lossless before $z_k$ and lossy after it, and the intensity peaks at $z_k$; for $Q_{\mathrm{st}} = -1$, the sequence is reversed and the intensity dips. The full-wave curves agree closely with the coupled-mode prediction of Eq. (9) (Fig. S2). The curvature at $z_k$ has the predicted sign for both windings and is only slightly smaller in magnitude, mainly because of couplings beyond nearest neighbors, which the coupled-mode model neglects but which do not change the sign of the winding.

Two controls without winding show no extremum. In the first, we use $g(z) = \tanh^2[(z - z_k)/\tau]$, so that $Q_{\mathrm{st}} = 0$. In this case, the loss imbalance vanishes at $z_k$ but recovers without changing sign (see Fig. S2). In the second control, all guides carry the same uniform loss. The comparison with the uniform control is particularly telling. Because $g(z)$ is odd about $z_k$ for $Q_{\mathrm{st}} = \pm 1$, the bound-state guides lose, on average, as much light as with uniform loss, and the three runs end with the same output power. The extremum is therefore set by the orientation of the switch, not by how much loss the light experiences.

This design also suggests a direct experiment. It requires only a passive sample and an intensity measurement along $z$, as routinely obtained by fluorescence imaging of the propagating light from above the sample[35]. Because $z_k$ lies at the center of the sample, launching light from the opposite facet replaces $g(z)$ with $-g(z)$ and reverses the winding, so both signs of $Q_{\mathrm{st}}$ can be compared in the same structure. The experiment is also tolerant to fabrication imperfections. By Eq. (5), only the sign change of the loss imbalance matters, not the precise shape of $g(z)$: a smoother or steeper transfer, or a slightly different loss level, changes the height of the extremum but not its type, and an error in $z_k$ merely shifts the extremum along with it.

## 4- Conclusion

These results show that topology can control not only where a bound state lives, but also when it brightens or dims. A single object, the mass field in space and time, does both. Its real part binds the state at the domain wall, and the windings of its zeros fix the timing of the response. One might have expected these two roles to require two separate mechanisms, a static interface for the confinement and an independent switch for the dynamics. The nonseparable fields of Fig. 3(a) show that this is not the case: when the wall moves and the switch is tilted, neither mechanism alone predicts the response, whereas the winding of the full mass field does.

The mechanism differs from temporal topology based on momentum gaps, in which a time-periodic modulation opens a gap in momentum and the associated states grow exponentially in time. Here, the bulk spectrum remains real and gapped throughout, no momentum gap opens, and the gain and loss act only through the sublattice polarization of a state that is already trapped. As a result, the effect requires no amplifying bulk and can be realized with loss alone. It also differs from the static complex SSH lattices in which staggered gain and loss select or amplify an interface mode: there, the gain-loss pattern decides which mode survives, whereas here its orientation in time decides when the selected mode peaks.

These properties suggest several uses in integrated photonics. Since the type of each extremum is fixed by an integer, a pattern of zeros written into a passive array acts as a robust program for the brightness of a confined mode: a single switch gives one peak or one dip, while a reversed region such as the one of Fig. 3(b) gives a peak, a dip and a second peak in a prescribed order. Such patterns could serve for gating and switching of light confined at a defect, or for shaping the intensity profile of a guided pulse along the propagation direction. In topological lasers based on SSH interface modes, the same principle could set the time at which the lasing mode experiences net gain. In this case, saturation and other nonlinear effects, absent from the present linear analysis, will need to be included.

Several extensions appear natural. In waveguide arrays, the switching pattern is fixed at fabrication, whereas time-multiplexed platforms, such as coupled fiber loops with controllable gain and loss, would allow it to be reconfigured from one experiment to the next. The analysis here assumes that the zeros are simple and well separated and that the modulation is slow compared with the bulk gap; relaxing these conditions, for instance by bringing two walls close enough for light to tunnel between them, may reveal new regimes in which the windings of neighboring zeros interact. Finally, in two and three dimensions, mass domain walls bind edge, surface and higher-order states[36,37]. An additional time-dependent mass that anticommutes with the binding mass could control their dynamics in the same way, opening a route to spacetime mass fields in which the position of a topological mode and the timing of its response are engineered independently.

# Supplementary Material

## S1- Mass field, winding, and response rule

Near the SSH band inversion, the slowly varying lattice is described by[1,2]

$$H_D(x,t) = \omega_0 I + iv\sigma_y\,\partial_x + m(x,t)\sigma_x + i\gamma(x,t)\sigma_z \qquad \text{(S1)},$$

where $\omega_0$ is the reference frequency, $v$ is the Dirac velocity, $m$ is the real mass set by the coupling dimerization, and $\gamma$ is the staggered imaginary mass (onsite terms $+i\gamma$ and $-i\gamma$ on sublattices A and B). The two masses multiply anticommuting matrices, and we track their joint direction through $M = m + i\gamma$.

Throughout this work, the imaginary mass stays well below the bulk gap. Let us consider a uniform lattice with intracell couplings $J_1$ and intercell couplings $J_2$. For the Bloch Hamiltonian $h(k) = J_1 + J_2 e^{-ik}$ and staggered term $i\gamma\sigma_z$, the bulk frequencies are

$$\omega(k) = \omega_0 \pm [|h(k)|^2 - \gamma^2]^{1/2} \qquad \text{(S2)}.$$

Since $\min_k|h(k)| = |J_1 - J_2| = 0.4$, while $|\gamma| \le \gamma_0 = 0.08$ in all calculations, the bulk spectrum is real for every $k$: no momentum gap opens, and the bulk remains in the same phase before and after $\gamma$ changes sign. In particular, the temporal switch does not create a temporal boundary state of the type of Ref.[3]. The only states that respond are those already bound by $m$.

At an isolated point $(x_d, t_k)$ where $m = \gamma = 0$, a counterclockwise contour $C$ with $M \neq 0$ on it defines the spacetime winding

$$Q_{\rm st} = \frac{1}{2\pi}\oint_C d\,\arg M \qquad \text{(S3)},$$

which counts how many times the direction of $(m,\gamma)$ turns around the origin along $C$. For a simple zero, the map $(x,t) \to (m,\gamma)$ is locally linear, and $Q_{\rm st}$ is the sign of its Jacobian,

$$Q_{\rm st} = \operatorname{sgn}\mathcal{J}, \qquad \mathcal{J} = \partial_x m\;\partial_t\gamma - \partial_t m\;\partial_x\gamma \quad \text{at } (x_d, t_k) \qquad \text{(S4)}.$$

For separable masses, $m = m(x)$ and $\gamma = \gamma(t)$, this reduces to $Q_{\rm st} = \operatorname{sgn}(\partial_x m\;\partial_t\gamma)$, so reversing either kink reverses $Q_{\rm st}$.

For $\gamma = 0$, the zero mode of Eq. (S1) satisfies $\partial_x\psi = (m/v)\sigma_z\psi$. It can therefore be chosen as an eigenstate of $\sigma_z$ with eigenvalue $s = \pm 1$ and envelope

$$\psi(x) \propto \exp\left[\frac{s}{v}\int_{x_d}^{x} m\,(x')\,dx'\right]\chi_s \qquad \text{(S5)}.$$

The state is normalizable only if the exponent decreases on both sides of the wall, which selects $s = -\operatorname{sgn}\partial_x m$. Reversing the wall therefore exchanges the occupied sublattice. In a finite chain, $s$ is evaluated as $\langle\psi|\Sigma_z|\psi\rangle$, where $\Sigma_z$ assigns $+1$ to A sites and $-1$ to B sites.

With the convention $i\dot{a} = \omega a$, the mode amplitude obeys $\dot{a} = s\gamma a$, and its power $P = |a|^2$ follows

$$\frac{d\ln P}{dt} = 2s\gamma(t) \qquad \text{(S6)}.$$

At the zero, $\gamma(t_k) = 0$ and $d^2\ln P/dt^2 = 2s\ \partial_t\gamma$. With $s = -\mathrm{sgn}\,\partial_x m$, this gives

$$\mathrm{sgn}\left[\frac{d^2\ln P}{dt^2}\right]\Bigg|_{t_k} = -Q_{\mathrm{st}} \qquad \text{(S7)}:$$

a maximum for $Q_{\mathrm{st}} = +1$ and a minimum for $Q_{\mathrm{st}} = -1$. For the control $\gamma = \gamma_0\tanh^2[(t - t_k)/\tau]$, $\gamma$ vanishes at $t_k$ but keeps its sign, so the image of a small contour does not encircle the origin and $Q_{\mathrm{st}} = 0$. Equation (S6) then has the same sign before and after $t_k$: $P$ becomes stationary at $t_k$ but remains monotonic, with no extremum.

When both masses depend on $x$ and $t$, the wall moves along $x_w(t)$, defined by $m(x_w, t) = 0$, with velocity $\dot{x}_w = -\,\partial_t m/\,\partial_x m$. A bound state that follows the wall adiabatically, and is narrow compared with the scale over which $\gamma$ varies, experiences the imaginary mass $\gamma_w(t) = \gamma(x_w(t), t)$, whose rate of change at the zero is

$$\frac{d\gamma_w}{dt} = \partial_t\gamma + \dot{x}_w\ \partial_x\gamma = \frac{\mathcal{J}}{\partial_x m} \qquad \text{(S8)}.$$

Equation (S6) with $\gamma \to \gamma_w$ then gives Eq. (S7) with the general winding of Eq. (S4).
For the profiles of Fig. 3(a), $m = m_0\tanh(X - aT)$ and $\gamma = \gamma_0\tanh(T - bX)$, with $X = (x - x_d)/\xi$ and $T = (t - t_k)/\tau$, where $\xi$ and $\tau$ set the widths of the two kinks in space and time. This gives $\mathcal{J} \propto (1 - ab)$ and $\gamma_w = \gamma_0\tanh[(1 - ab)T]$. The two cases compared there, $b = 2$ and $b = 6$ with $a = 0.25$, have $1 - ab = \pm 0.5$ and therefore opposite windings.

The phase $\theta = \arg M$ is smooth wherever $M \neq 0$. For a region R whose boundary $\partial$R avoids the zeros, the boundary winding

$$W_{\mathrm{R}} = \frac{1}{2\pi}\oint_{\partial \mathrm{R}} d\,\theta = \sum_{i\in R} Q_{\mathrm{st},i} \qquad \text{(S9)}$$

equals the sum of the windings of the zeros inside R. The reason is that $\oint d\theta$ vanishes around any loop enclosing no zero, so $\partial$R can be shrunk onto small circles around each zero. When the zeros are simple and well separated, so that each bound state extremum is set by a single zero, Eq. (S7) assigns a maximum to each $Q_{\mathrm{st}} = +1$ and a minimum to each $Q_{\mathrm{st}} = -1$. This gives $N_{\max} - N_{\min} = W_{\mathrm{R}}$.
Two zeros of opposite winding can merge only through a degenerate point where $\mathcal{J} = 0$. This is what happens in Fig. 3(c): when the $\gamma$-reversed region just touches the wall, the zero line of $\gamma$ becomes tangent to the zero line of $m$, the two zeros coincide with $\mathcal{J} = 0$, and they annihilate together with their extrema.

## S2- Finite-chain simulations and numerical parameters

We consider open chains of $N$ unit cells with index $j$, sites $A_j$ and $B_j$, intracell couplings $J_{1,j}$ and intercell couplings $J_{2,j}$, so that the local mass is $m_j = J_{1,j} - J_{2,j}$[4]. Frequencies are measured from $\omega_0$, and the time-dependent Hamiltonian is

$$H(t) = H_0 + i\Gamma(t), \Gamma(t) = \sum_j \gamma_j(t)\big(|A_j\rangle\langle A_j| - |B_j\rangle\langle B_j|\big) \qquad \text{(S10)},$$

where $H_0$ is the passive SSH Hamiltonian. The sublattice polarization of a normalized state is $s = \langle\psi|\Sigma_z|\psi\rangle$, with $\Sigma_z = +1$ on A sites and $-1$ on B sites. The passive boundary and interface states of a finite chain are nearly degenerate, so direct diagonalization can return mixed combinations of the two. We therefore diagonalize the position operator within the two-dimensional midgap subspace, which yields one state localized at the outer boundary and one at the domain wall. The single mode prediction is Eq. (S6) integrated in time: $P(t) = P(t_0)\exp\big[2s\int \gamma\, dt'\big]$.

The parameters used in Figs. 1 and 2 are the following; the chain has $N = 80$ cells, with $(J_1, J_2) = (0.8, 1.2)$ for $j \leq 40$ and $(1.2, 0.8)$ for $j > 40$, so that $m < 0$ on the left and $m > 0$ on the right. The two midgap states have $s_b \simeq +1$ and $s_{\text{int}} \simeq -1$. For Fig. 1(c), the chain starts empty, and a Gaussian source $F(t) = F_0\exp[-(t - t_s)^2/(2\sigma_s^2)]$, with $F_0 = 0.03$, $t_s = 12$ and $\sigma_s = 6$, drives the site with the largest weight of the interface state, well before the switch. The intensity map is normalized by its global maximum. For Fig. 2(b), we diagonalize $H_0 + i\gamma\Sigma_z$ for 125 values of $|\gamma|$ between 0.002 and 0.25 for each sign, and follow each localized branch by its overlap with the passive state. For Figs. 2(c) and 2(d), the evolution starts directly in the passive interface state, and the power is normalized to its value at $t_k$. In all cases, $\gamma(t) = \pm\gamma_0\tanh[(t - t_k)/\tau]$ or $\gamma_0\tanh^2[(t - t_k)/\tau]$, with $\gamma_0 = 0.08$, $t_k = 60$ and $\tau = 3$.

The parameters used in Fig. 3(a) are the following; the chain has $N = 80$ cells, with couplings $J_{1,2} = 1 \pm m_j(t)/2$ reconstructed at each time from the profiles of Sec. 1. The parameters are $x_d = 40.5$, $\xi = 12$, $t_k = 60$, $\tau = 15$, $m_0 = 0.40$, $\gamma_0 = 0.04$, $a = 0.25$, and $b = 2$ or 6. Because the wall moves, the interface state is re-identified at every time step from the instantaneous passive Hamiltonian, and its power is $P(t) = |\langle\phi_{\text{int}}(t) \mid \psi(t)\rangle|^2$.

The parameters used in Figs. 3(b) and 3(c) are the following; the chain has $N = 100$ cells, with domain walls after cells 35 and 65: $(J_1, J_2) = (1.2, 0.8)$ in the outer regions and $(0.8, 1.2)$ in between. Under the common switch, the left wall therefore carries $Q_{\text{st}} = -1$ and the right wall $Q_{\text{st}} = +1$. The imaginary mass is

$$\gamma_j(t) = \gamma_0 \tanh\left(\frac{t}{\tau}\right) \tanh\left\{\frac{\left[\frac{(j-x_b)}{R_x}\right]^2 + \left[\frac{(t-t_b)}{R_t}\right]^2 - 1}{d}\right\} \text{(S11)},$$

with $\gamma_0 = 0.08$, $\tau = 3$ and $t_k = 0$. The second factor equals $-1$ inside an ellipse of half-widths $R_x = 7$ cells and $R_t = 4$, centered at $x_b = 65.5 + \Delta x$ and $t_b = 10$, and $+1$ outside; $d = 0.25$ sets the sharpness of its edge. The evolution starts in the equal superposition of the two interface states, and the power of each is obtained by projection. The boundary of the ellipse crosses the right wall at

$$t - t_k = t_b \pm R_t \left[1 - \left(\frac{\Delta x}{R_x}\right)^2\right]^{1/2} \text{(S12)},$$

for $|\Delta x| < R_x$. These are the predicted zeros shown as lines in Fig. 3(c). Fig. 3(b) uses $\Delta x = 3$.

The counting rule assumes that each bound state responds independently. When two walls approach within a few localization lengths of each other, their states overlap and light can tunnel between them; once this tunnelling becomes comparable to the gain and loss, it dominates the dynamics and the extrema are no longer fixed by the windings. In all configurations of the main text, the walls are separated by many localization lengths (at least 30 unit cells, compared with a localization length of about 2.5 cells).

## S3- Disorder

The clean chains of Secs. 1 and 2 isolate the relation between winding and response. To test its robustness, we add independent disorder to the couplings and to the local amplitude of the staggered imaginary mass in the chain of Figs. 1 and 2:

$$H^{(\text{dis})}(t) = H_0^{(\text{dis})} + i\gamma(t) \sum_j \left[\left(1 + \zeta_{A,j}\right)|A_j\rangle\langle A_j| - \left(1 + \zeta_{B,j}\right)|B_j\rangle\langle B_j|\right] \text{(S13)},$$

where $H_0^{(\text{dis})}$ has couplings $J_{1,j}(1 + \eta_{1,j})$ and $J_{2,j}(1 + \eta_{2,j})$. The variables $\eta$ and $\zeta$ are independent and uniformly distributed in $[-V, V]$, where $V$ is the maximum relative variation of each coupling and each gain/loss amplitude. We use $V = 0.30$, so each coupling and each local gain/loss amplitude varies by up to 30%. Over this range, the gain and loss keep opposite signs on the two sublattices, and the average bulk gap, $|J_1 - J_2| = 0.4$, remains well above $|\gamma|$, although individual couplings can locally reverse the dimerization.

For each of 50 realizations, the boundary and interface states are re-identified as in Sec. 2. Each is then evolved under $\gamma(t) = \gamma_0 \tanh[(t - t_k)/\tau]$, with its power normalized to its initial value. Fig. S1 shows the ensemble averages: the interface state keeps its growth then decay sequence, and the boundary state keeps the opposite one. Disorder changes the size of the response and broadens its spread, but the opposite temporal orientations remain clear.

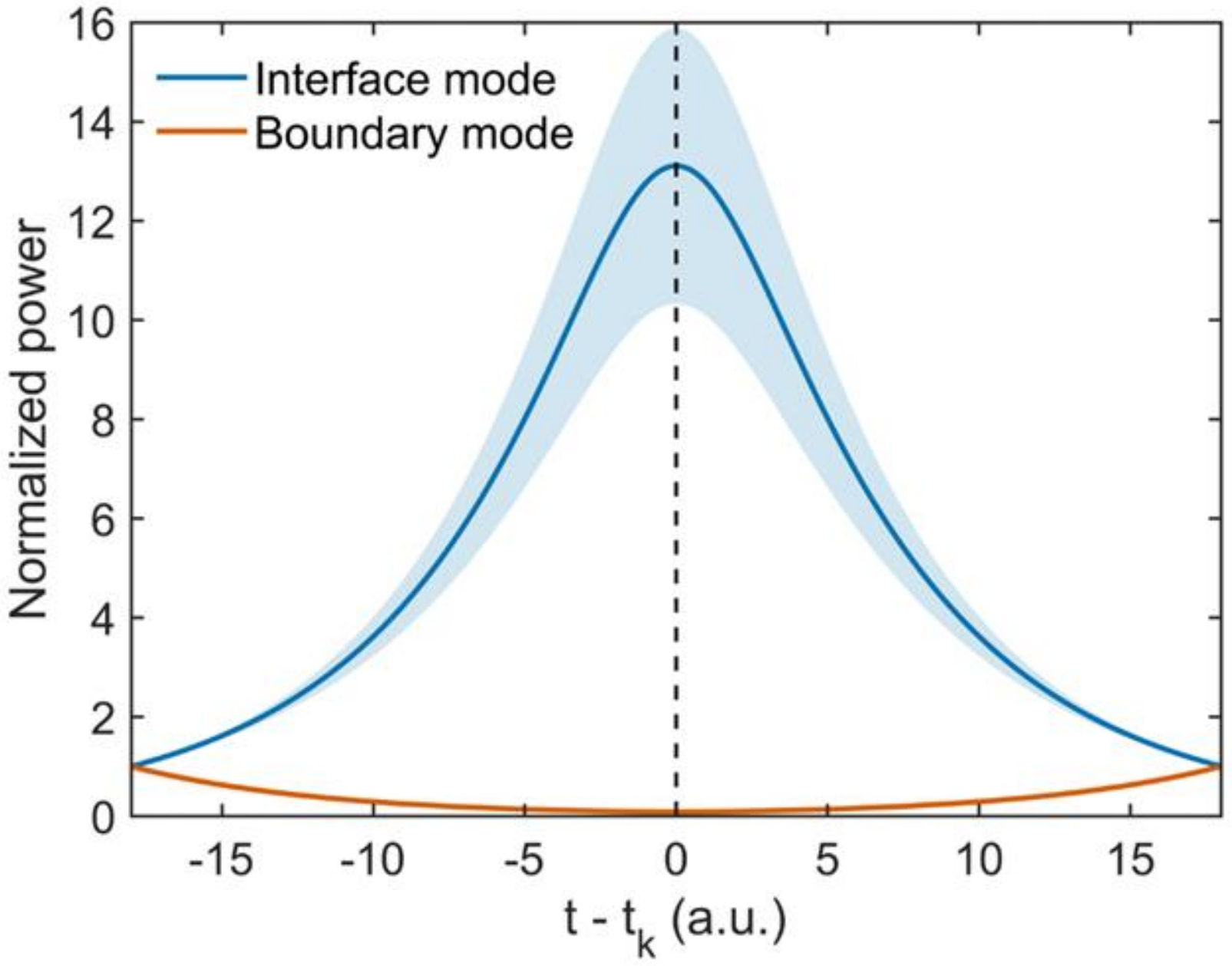

**Figure S1.** Ensemble-averaged normalized powers of the interface and boundary states for 50 realizations of simultaneous coupling and gain/loss disorder ($V = 0.30$). Shaded regions: one standard deviation. Dashed line: $t = t_k$.

## S4- Full-wave simulations of the waveguide array

We model a two-dimensional $(x, z)$ array of 60 slab waveguides of width 4 μm, with refractive index $n_{\mathrm{clad}} + \Delta n$, where $n_{\mathrm{clad}} = 1.457$ and $\Delta n = 1.2 \times 10^{-3}$, at a wavelength $\lambda = 633$ nm. Each guide supports a single mode. The center-to-center gaps alternate between 8.0 μm and 8.9 μm, giving strong and weak couplings $J_s = 0.60\ \mathrm{mm}^{-1}$ and $J_w = 0.40\ \mathrm{mm}^{-1}$, obtained from the splitting of the two-guide supermodes. On the left of the wall, the weak gap lies inside each unit cell; on the right, the strong gap does. The central A guide (see Fig.4a of the main text) lies between two strong gaps and forms the domain wall. The bound state occupies only the B guides, with amplitudes that alternate in sign and decrease by a factor $J_w/J_s = 2/3$ per unit cell away from the wall. At $z = 0$, we launch this state as a sum of Gaussians centered on the B guides, weighted by these amplitudes, with a width chosen to match the single guide mode.

Loss enters through the imaginary part $k \geq 0$ of the guide index, $n = n_r - ik$, with

$$k_A = \frac{\kappa[1 - g(z)]}{2}, \qquad k_B = \frac{\kappa[1 + g(z)]}{2} \qquad \text{(S14)}.$$

Here $g(z) = \pm\tanh[(z - z_k)/\tau]$ for $Q_{\mathrm{st}} = \pm 1$, $g(z) = \tanh^2[(z - z_k)/\tau]$ for the zero winding control, and $g = 0$ for uniform loss. We use $z_k = 30$ mm, $\tau = 6$ mm, and a sample length $L = 60$ mm. The constant $\kappa$ is chosen so that a fully lossy guide loses 0.43 dB/mm in power, which corresponds to a mean loss rate $\Gamma = 0.025\ \mathrm{mm}^{-1}$ for the array. This level is comparable to that obtained by bending femtosecond-written waveguide arrays[5]. In coupled-mode form, the guide amplitudes obey

$$i\,\frac{d\psi}{dz} = [H - i\Gamma + i\Gamma g(z)\,\sigma_z]\,\psi \qquad \text{(S15)},$$

where $H$ contains the passive couplings and $\sigma_z = +1$ on A guides and $-1$ on B guides. This is the staggered term of Eq. (S1), with $z$ in place of $t$ and $\gamma = \Gamma g$, together with a uniform loss $\Gamma$. The uniform term multiplies every amplitude by $e^{-\Gamma z}$. It therefore adds a linear term to $\ln P$ and leaves its curvature unchanged, which is why we display $Pe^{2\Gamma z}$. For the bound state, which lives on the B guides, Eqs. (S6) and (S15) give

$$\frac{d\ln P}{dz} = -2\Gamma\left[1 \pm \tanh\left(\frac{z - z_k}{\tau}\right)\right] \qquad \text{(S16)}.$$

For $Q_{\mathrm{st}} = +1$, the bound state is lossless before $z_k$ and lossy after it, which produces a maximum of $Pe^{2\Gamma z}$ at $z_k$. For $Q_{\mathrm{st}} = -1$, the sequence is reversed and a minimum appears. Because $z_k$ lies at the center of the sample, launching light from the opposite facet replaces $g(z)$ with $-g(z)$, so the same structure gives the opposite winding.

We solve Maxwell's equations with the beam-envelope method of COMSOL Multiphysics, using a unidirectional envelope along $z$ and an out-of-plane polarization. The input and output faces carry matched boundary conditions, and the lateral boundaries are scattering boundaries. Without loss, the launched bound state propagates without changing shape. With uniform loss, $\ln P$ decreases linearly at the expected rate $2\Gamma$. The launched profile reproduces the target amplitudes closely.

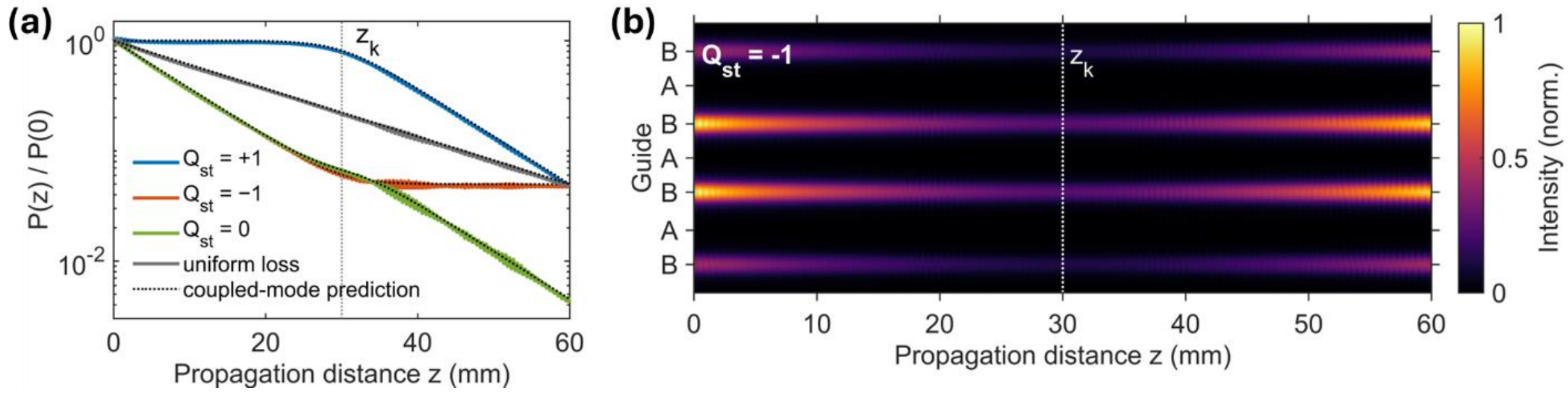

**Figure S2.** Full-wave results for the waveguide array. **(a)** Intensity at the center of the bound-state guide for $Q_{\mathrm{st}} = +1$, $Q_{\mathrm{st}} = -1$, the zero winding control and uniform loss (solid lines), compared with the coupled-mode prediction (dotted lines). **(b)** Intensity $|E|^2 e^{2\Gamma z}$ for $Q_{\mathrm{st}} = -1$.

We record the intensity $P(z)$ at the center of a B guide next to the wall. The full-wave curves follow the coupled-mode prediction of Eq. (S16) closely [Fig. S2(a)]. The curvature at $z_k$ has the predicted sign for both windings and is slightly smaller in magnitude, which we attribute mainly to couplings beyond nearest neighbours, absent from the coupled-mode model. The two controls show no curvature at $z_k$. The runs with $Q_{\rm st} = \pm 1$ and with uniform loss end with the same output power, so they remove the same total amount of light, while the zero winding control, which keeps the bound-state guides lossy on both sides of $z_k$, ends lower. Figure S2(b) shows the intensity map for $Q_{\rm st} = -1$, the counterpart of Fig. 4(b): as for $Q_{\rm st} = +1$, the light stays on the B guides throughout.